\pdfoutput=1
\documentclass{article}
\usepackage{dcase2021,amsmath,graphicx,url,times,booktabs, tabularx, hyperref}
\usepackage{multirow}

\newcommand\blfootnote[1]{%
  \begingroup
  \renewcommand\thefootnote{}\footnote{#1}%
  \addtocounter{footnote}{-1}%
  \endgroup
}

\usepackage{tikz}
\def\checkmark{\tikz\fill[scale=0.4](0,.35) -- (.25,0) -- (1,.7) -- (.25,.15) -- cycle;} 

\title{The impact of non-target events in synthetic soundscapes for sound event detection}
\name{Francesca Ronchini$^{1}$,
       Romain Serizel$^{1}$,
       Nicolas Turpault$^{1}$, 
       Samuele Cornell$^{2}$
       }
\address{$^1$Université de Lorraine, CNRS, Inria, Loria, Nancy, France \\
         $^2$ Department of Information Engineering, Universita Politecnica delle Marche, Italy \\
  }

\begin{document}

\ninept
\maketitle

\begin{sloppy}

\begin{abstract}
Detection and Classification Acoustic Scene and Events Challenge 2021 Task 4 uses a heterogeneous dataset that includes both recorded and synthetic soundscapes. Until recently only target sound events were considered when synthesizing the soundscapes. However, recorded soundscapes often contain a substantial amount of non-target events that may affect the performance. In this paper, we focus on the impact of these non-target events in the synthetic soundscapes. Firstly, we investigate to what extent using non-target events alternatively during the training or validation phase (or none of them) helps the system to correctly detect target events. Secondly, we analyze to what extend adjusting the signal-to-noise ratio between target and non-target events at training improves the sound event detection performance. 
The results show that using both target and non-target events for only one of the phases (validation or training) helps the system to properly detect sound events, outperforming the baseline (which uses non-target events in both phases).
The paper also reports the results of a preliminary study on evaluating the system on clips that contain only non-target events. This opens questions for future work on non-target subset and acoustic similarity between target and non-target events which might confuse the system. 
\end{abstract}

\begin{keywords}
Sound event detection, synthetic soundscapes, open-source datasets, deep learning
\end{keywords}

\section{Introduction}
\label{sec:intro}

\blfootnote{This work was made with the support of the French National Research Agency, in the framework of the project LEAUDS Learning to understand audio scenes (ANR-18-CE23-0020), the project CPS4EU Cyber Physical Systems for Europe (Grant Agreement number: 826276) and the French region Grand-Est. Experiments presented in this paper were carried out using the Grid5000 testbed, supported by a scientific interest group hosted by Inria and including CNRS, RENATER and several Universities as well as other organizations (see \href{https://www.grid5000}{https://www.grid5000}).}

The main goal of ambient sound and scene analysis is to automatically extract information from sounds that surround us and analyze them for different purposes and applications. 
Between the different area of interest, ambient sound analysis have a considerable impact on applications such as noise monitoring in smart cities \cite{bello2018sonyc, bello2018sound}, domestic applications such as smart homes and home security solutions \cite{serizel2018large, debes2016monitoring}, health monitoring systems \cite{zigel2009method}, multimedia information retrieval \cite{jin2012event} and bioacoustics domain \cite{morfi2021deep}. 
Sound Event Detection (SED) aims to identify the onset and offset of the sound events present in a soundscape and to correctly classify them, labeling the events according to the target sound classes that they belong to. Nowadays, deep learning is the main method used to approach the problem. However, one of the main limitations of deep learning models is the requirement of large amounts of labeled training data to reach good performance. The process of labeling data is time-consuming and bias-prone mainly due to human errors and disagreement given the subjectivity in the perception of some sound event onsets and offsets \cite{Turpault2019_DCASE}. To overcome these limitations, recent works are investigating alternatives to train deep neural networks with a small amount of labeled data together with a bigger set of unlabeled data \cite{serizel2018large, serizel2019sound, shah2018closer, Turpault2019_DCASE, mcfee2018adaptive}. Among them, Detection and Classification Acoustic Scenes and Events Challenge (DCASE) 2021 Task 4 uses an heterogeneous dataset that includes both recorded and synthetic soundscapes \cite{Turpault2019_DCASE}. 
This latter soundscapes provide a cheap way to obtain strongly labeled data. Until recently, synthesized soundscapes were generated considering only target sound events. However, recorded soundscapes also contain a considerable amount of non-target events that might influence the performance of the system. 

The purpose of this paper is to focus on the impact on the system's performance when non-target events are included in the synthetic soundscapes of the training dataset. The study has been mainly divided into three stages.
Firstly, we investigate to what extent using non-target events alternatively during training or validation helps the system to correctly detect the target sound events. Mainly motivated from the results of the first experiment, in the second part of the study, we focus on understanding to what extend adjusting the target to non-target signal-to-noise ratio (TNTSNR) at training improves the sound event detection performance. Results regarding a preliminary study on the evaluation of the system using clips containing only non-target events are also reported, opening questions for future studies on possible acoustic similarity between target and non-target sound events which might confuse the SED system. \footnote{To promote reproducibility, the code, \url{https://github.com/DCASE-REPO/DESED_task},
and pre-trained models \url{https://zenodo.org/record/5529692}, are made available under an open-source license.}.



\section{PROBLEM DEFINITION AND DATASET GENERATION}
\label{sec:format}

\subsection{Problem definition}
\label{ssec:probeldef}
The primary goal of the DCASE 2021 Challenge Task 4 is the development of a semi-supervised system for SED, exploiting an heterogeneous and unbalanced training dataset. 
The goal of the system is to correctly classify the sound event classes and to localize the different target sound events present in an audio clip in terms of timing. Each audio recording can contain more than one event. Some of those could also be overlapped. 
The use of a larger amount of unlabeled recorded clips is motivated by the limitations related to annotating a SED dataset (human-error-prone and time-consuming). 
Alternatively, synthesized soundscapes are an easy way to have strongly annotated data. In fact, the user can easily generate the soundscapes starting from isolated sound events. 
On the other hand, in most of the recorded soundscapes the target sound classes are almost never present alone. 
For this reason, one of the main novelties of the DCASE 2021 Challenge Task 4 is the introduction of non-target isolated events in the synthetic soundscapes\footnote{\url{http://dcase.community/challenge2021}}.
 This paper explores the impact of the non-target sound events on the baseline system performance, with the final goal of understanding and highlighting how to correctly exploit them to generate realistic soundscapes.
 
\subsection{Dataset generation}
\label{ssec:datasetgen}
The dataset used in this paper is the DESED dataset\footnote{\href{https://project.inria.fr/desed/}{https://project.inria.fr/desed/}} \cite{serizel:hal-02355573, turpault:hal-02160855}, which is the same provided for the DCASE 2021 Challenge Task 4. 
It is composed of 10 seconds length audio clips either recorded in a domestic environment or synthesized to reproduce such an environment\footnote{For a detailed description of the DESED dataset and how it is generated the reader is referred to the original DESED article~\cite{turpault:hal-02160855} and DCASE 2021 task 4 webpage: \url{http://dcase.community/challenge2021}}.   
The synthetic part of the dataset is generated with Scaper \cite{salamon2017scaper}, a Python library for soundscape synthesis and augmentation, 
which allows to control audio parameters. 
The recorded soundscapes are taken from AudioSet \cite{gemmeke2017audio}. The foreground events (both target and non-target) are obtained from the Freesound Dataset (FSD50k) \cite{fonseca2020fsd50k}, while the background sounds are obtained from the SINS dataset (activity class “other”) \cite{dekkers2017sins} and TUT scenes 2016 development dataset \cite{mesaros2016tut}. In particular, non-target events are the intersection of FUSS dataset \cite{wisdom2021s} and FSD50k dataset in order to have compatibilty with the source separation baseline system.
 
In this article, we modify only the synthetic subset of the dataset. Starting from the synthetic part of the DESED dataset, we generated different versions of it in order to investigate how non-target events impact the system performance and to what extent their relationship with the target events affects the training phase of the system. 
The following subsections describe the different subsets used for the experiments, which have been generated using Scaper. 

\subsubsection{Synthetic training set}
\label{sssec:trainset}
The synthetic training set is the same set of data released for the DCASE 2021 Challenge Task 4. It includes 10000 audio clips where both target and non-target sound events could be present in each clip. The distribution of the sound events among the files have been determined considering the co-occurrences between the different sound events. The co-occurrences have been calculate considering the strong annotations released for the AudioSet dataset~\cite{hershey2021benefit}\footnote{The co-occurrences distribution and the code used to compute them will be distributed.}. A second version of this dataset has been generated where only target events are present. 
The datasets will be hereafter referred as \textbf{synth\_tg\_ntg} (used by the official baseline system) and \textbf{synth\_tg} for the synthetic subset including target and non-target events and the synthetic subset including only target events, respectively. 

\subsubsection{Synthetic validation set}
\label{sssec:valset}
The synthetic validation set is the same as the synthetic validation dataset supplied for the DCASE 2021 Challenge Task 4. It includes 3000 audio clips including target and non-target events, which distribution has been defined calculating the co-occurrences between sound events. We generated a second version of the dataset containing only target events. The datasets will be referred to as \textbf{synth\_tg\_ntg\_val} (used by the baseline system) and \textbf{synth\_tg\_val} (only target sound events). 

\subsubsection{Synthetic evaluation set}
\label{sssec:synth21eval}
The synthetic 2021 evaluation set is composed by 1000 audio clips. 
In the context of the challenge, this subset is used for analysis purposes. We will refer to it as \textbf{synth\_tg\_ntg\_eval}. It contains target and non-target events distributed between the different audio clips according to the pre-calculated co-occurrences. Two different versions of the \textbf{synth\_tg\_ntg\_eval} set have been generated, \textbf{synth\_tg\_eval} (only target sound events) and \textbf{synth\_ntg\_eval} (only non-target sound events). 

\subsubsection{Varying TNTSNR training and validation set}
\label{sssec:TNTSNRset}
With the aim of studying what would be the impact of varying the TNTSNR on the system performance, different versions of \textbf{synth\_tg\_ntg} and \textbf{synth\_tg\_ntg\_val} have been generated. In particular, for each of them, three versions have been created. The SNR of the non-target events have been decreased by 5 dB, 10 dB and 15 dB compared to their original value. The original SNR of the sound events is randomly selected between 6 dB and 30 dB, so the more we decrease the SNR, the less the sound will be audible, with some of the events that will not be audible at all. These subsets will be subsequently referred to as \textbf{synth\_5dB}, \textbf{synth\_10dB}, \textbf{synth\_15dB} for the training subsets and \textbf{synth\_5dB\_val}, \textbf{synth\_10dB\_val}, \textbf{synth\_15dB\_val} for the validation subsets. 

\subsubsection{Public evaluation set}
\label{sssec:publicset}
The public evaluation set is composed of recorded audio clips extracted from Youtube videos that are under creative common licenses. 
This is part of the evaluation dataset released for the evaluation phase of the DCASE 2021 Challenge Task 4 and considered for ranking. 
The set will be referred to as \textbf{public}.

\section{EXPERIMENTS TASK SETUP}
\label{sec:experiments}

In order to compare the results with the official baseline, we used the same SED
mean-teacher system released for this year challenge. More information regarding the system can be found at Turpault et al.~\cite{Turpault2019_DCASE} and on the official webpage of the DCASE Challenge Task 4. 
All the different models have been trained 5 times. This paper reports the average of the scores and the confidence intervals related to those. Only for the baseline model we do no report the confidence intervals because we have considered the results using the checkpoint made available for it \footnote{\url{https://zenodo.org/record/4639817}}. The metrics considered for the study are the two polyphonic sound detection score (PSDS)~\cite{bilen2020framework} scenarios defined for the DCASE 2021 Challenge Task 4, since these are the official metrics used in the challenge.

The scope of these experiments is twofold: understand the impact of non-target events on the system performance and investigate to what extend the TNTSNR helps the network to correctly predict the sound events in both matched and mismatched conditions. In order to do so, we divided the experiment into three stages. The first part of the study is focused on understanding the influence of training the system with non-target events. This experiment is described and discussed in Section ~\ref{sec:tgntgexp}. Section ~\ref{sec:varyingTNT} reports the results and the relative discussion of the second part of the experiment where we investigate if a mismatch in terms of TNTSNR between datasets could have an impact on the output of the system. Section ~\ref{sec:notgeval} reports preliminary results of the last stage of the experiment, regarding the evaluation of the system on the \textbf{synth\_ntg\_eval} dataset, formed by only non-target sound events, in order to investigate if some classes could get acoustically confused at training, having a negative impact on the performance. The last stage has been motivated by the results of the second part of the experiment.

\section{Using TARGET/NON-TARGET at training}
\label{sec:tgntgexp}

\begin{table}[t]
\centering
 \begin{tabular}{c|c|c|c}
 \toprule
  \multicolumn{2}{c|}{\textbf{Non-target}} & \multirow{2}{*}{\textbf{PSDS1}} & \multirow{2}{*}{\textbf{PSDS2}} \\
  \cline{0-1}
  Train & Val&& \\
  \midrule
  \checkmark &  & 33.81 (0.36) & 52.62 (0.19) \\
  & \checkmark & 35.92 (0.49) & 54.85 (0.29) \\
  & & 34.90 (0.82) & 53.07 (1.22) \\
  \checkmark & \checkmark & \textbf{36.40} & \textbf{58.00} \\
 \bottomrule
 \end{tabular}
 \caption{Evaluation results for the \textbf{public} set, considering the different combinations of using target and non-target sound events at training and validation.}
 \label{tab:results_public}
\end{table} 

In the first experiment we concentrate on training the system with different combinations of the training dataset. 
Table ~\ref{tab:results_public} reports the results of the experiment evaluating the system on the \textbf{public} set. We check-marked the columns NT Train or/and NT Val according to if the non-target sound events are present or not in the synthetic sounscapes. From the results it is possible to observe that using non-target sound events  during training and validation improves the performance by a large margin with relaxed segmentation constraints (PSDS2) but only marginally with strict segmentation constraints (PSDS1). In this latter case what matters the most is the use of non-target sound events during the validation. A possible explanation is that synthetic soundscapes with non-target sound events are actually too difficult and confuse the systems when used during the training but they still help reducing the mismatch with recorded soundscapes during model selection (validation).

\begin{table}[!t]
 \begin{tabularx}{\columnwidth}{c@{\hskip0.10in}|c@{\hskip0.10in}|c@{\hskip 0.10in}|c@{\hskip0.10in}|c@{\hskip0.10in}}
  \toprule
  \multicolumn{2}{c|}{\textbf{Non-target}} & \multirow{2}*{\textbf{Eval set}} & \multirow{2}*{\textbf{PSDS1}} & \multirow{2}*{\textbf{PSDS2}} \\
  \cline{0-1}
  Train & Val&& \\
  \midrule 
  \checkmark &  & synth\_tg\_ntg\_eval & 23.22 (1.33) & 36.44 (2.62) \\
  & \checkmark & synth\_tg\_ntg\_eval &  20.08 (0.39) & 31.33 (1.29) \\
  & & synth\_tg\_ntg\_eval & 20.13 (0.35) & 30.99 (1.07) \\
  \checkmark & \checkmark & synth\_tg\_ntg\_eval & \textbf{25.14} & \textbf{40.12} \\
  \hline
  \hline
  \checkmark &  & synth\_tg\_eval & 42.82 (2.42) & 58.26 (2.08) \\
   & \checkmark & synth\_tg\_eval &  46.92 (1.02) & \textbf{62.79 (0.55)} \\
   & & synth\_tg\_eval  & \textbf{47.73 (0.33)} & 62.54 (1.00) \\
  \checkmark & \checkmark & synth\_tg\_eval & 43.22 & 61.09 \\
 \bottomrule
 \end{tabularx}
 \caption{Evaluation results for the \textbf{synth\_tg\_ntg\_eval} set and \textbf{synth\_tg\_eval} set, considering the different combination of using target and non-target sound events at training and validation.}
 \label{tab:results_tgeval}
\end{table} 


Table ~\ref{tab:results_tgeval} reports the results considering the \textbf{synth\_tg\_ntg\_eval} and \textbf{synth\_tg\_eval} evaluation sets. In all cases the best performance is obtained in matched training/evaluation conditions. The performance obtained on \textbf{synth\_tg\_ntg\_eval} are lower than the performance obtained on \textbf{synth\_tg\_eval} even in matched conditions. Not surprisingly, this confirm that including non-target sound events makes the SED task more difficult. Interestingly, as opposed to the previous experiment, the most important here is to have matched conditions during training and to a lesser extent during validation. In order to verify the low impact of non-target sound events at training when evaluating on recorded soundscapes, in the next experiment we investigate a possible mismatch in terms in TNTSNR.

\section{Varying TNTSNR at training}
\label{sec:varyingTNT}

\begin{table}[!t]
\centering
 \begin{tabular}{c|c|c|c}
 \toprule
  \multicolumn{2}{c|}{\textbf{Non-target}} & \multirow{2}*{\textbf{PSDS1}} & \multirow{2}*{\textbf{PSDS2}} \\
  \cline{0-1}
  Train & Val&& \\
 \midrule 
  Original & 5~dB & 35.57 (0.28) & 56.68 (1.77) \\
  5~dB & Original & 36.25 (1.26) & 57.53 (1.06) \\
  5~dB & 5~dB & 35.46 (0.46) & \textbf{58.09} (0.74) \\
  Original & Original & \textbf{36.40} & 58.00 \\
 \bottomrule
 \end{tabular}
 \caption{Evaluation results for the second part of the experiment, varying TNTSNR by 5 dB (\textbf{synth\_5dB} and \textbf{synth\_5dB\_val}). Evaluating with \textbf{public} set.}
 \label{tab:results_5db}
\end{table} 

\begin{table}[!t]
  \centering
 \begin{tabular}{c|c|c|c}
 \toprule
  \multicolumn{2}{c|}{\textbf{Non-target}} & \multirow{2}*{\textbf{PSDS1}} & \multirow{2}*{\textbf{PSDS2}} \\
  \cline{0-1}
  Train & Val &&\\
 \midrule 
  Original & 10~db & 36.23 (1.11) & 57.82 (1.37) \\
  10~db & Original & \textbf{36.42 (0.77)} & \textbf{58.94 (0.89)} \\
  10~db & 10~db & 36.20 (1.14) & 57.92 (1.04) \\
  Original & Original & 36.40 & 58.00 \\
 \bottomrule
 \end{tabular}
 \caption{Evaluation results for the second part of the experiment, varying TNTSNR by 10 dB (\textbf{synth\_10dB} and \textbf{synth\_10dB\_val}). Evaluating with \textbf{public} set.}
 \label{tab:results_10db}
\end{table} 

\begin{table}[!t]
\centering
 \begin{tabular}{c|c|c|c} 
 \toprule
  \multicolumn{2}{c|}{\textbf{Non-target}} & \multirow{2}*{\textbf{PSDS1}} & \multirow{2}*{\textbf{PSDS2}} \\
  \cline{0-1}
  Train & Val \\
 \midrule 
  Original & 15~dB & 36.08 (1.13) & 57.78 (1.33) \\
  15~dB & Original & \textbf{37.37 (0.70)} & \textbf{58.64 (1.34)} \\
  15~dB & 15~dB & 36.10 (0.50) & 57.36 (0.89) \\
  Original & Original & 36.40 & 58.00 \\
 \bottomrule
 \end{tabular}
 \caption{Evaluation results for the second part of the experiment, varying TNTSNR by 15 dB (\textbf{synth\_15dB} and \textbf{synth\_15dB\_val}). Evaluating with \textbf{public} set.}
 \label{tab:results_15db}
\end{table}

\begin{table}[!t]
\centering
 \begin{tabular}{c|c|c}
  \toprule
  Validation set & PSDS1 & PSDS2 \\
  \midrule 
  synth\_5dB\_val & 38.68 (1.07) & 60.57 (0.78) \\
  synth\_10dB\_val & \textbf{39.07 (0.75)} & \textbf{60.75 (0.80)} \\
  synth\_15dB\_val & 37.95 (0.53) & 59.99 (1.14) \\
 \bottomrule
 \end{tabular}
 \caption{Evaluation results of the SED system, training with \textbf{synth\_tg}, validating with varying TNTNSNR set and evaluating with \textbf{public} set.}
 \label{tab:results_tgSNR}
\end{table} 

The second part of the study focuses on understanding the impact of varying the TNTSNR at training and validation aiming at finding a TNTSNR condition that could match better the recorded soundscapes. 
For each TNTSNR, we use similar combinations 
as the ones used in Section ~\ref{sec:tgntgexp}, replacing the set without non-target sound events by a set with adjusted TNTSNR.
For example, considering the 5 dB case, the combinations considered would be: 
\begin{itemize}
  \item training using the \textbf{synth\_tg\_ntg} set and validating with \textbf{synth\_5dB\_val}; 
  \item training with \textbf{synth\_5dB} and validating with \textbf{synth\_tg\_ntg\_val}; 
  \item training and validating with \textbf{synth\_5dB} and \textbf{synth\_5dB\_val}. 
\end{itemize}
The fourth combination is the official DCASE Task 4 baseline. 
Repeating the experiment with all the varying TNTSNR, allow us to analyse to what extend the loudness of the non-target events helps matching the evaluation conditions on recorded clips. 
Table ~\ref{tab:results_5db}, ~\ref{tab:results_10db} and ~\ref{tab:results_15db} report the performance on the \textbf{public} set when using a TNTSNR of 5~dB, 10~dB and 15~dB, respectively. When the TNTSNR is 5~dB or 10~dB, the performance changes only marginally between configurations. Increasing the TNTSNR to 15~dB leads to a behaviour more similar to the one obtained in Table~\ref{tab:results_public}. The best performance is obtained when training with TNTSNR is 15~dB and validating on \textbf{synth\_tg\_ntg\_val}. This could be explained by the fact TNTSNR 15~dB is a condition closer to that of the recorded soundscapes and the fact that it allows for selecting models that will be more robust towards non-target events at test time.

In the last experiment, we investigate the impact of varying the TNTSNR during validation phase, while using the \textbf{synt\_tg} for training. 
Results are reported on Table \ref{tab:results_tgSNR}, where it is possible to observe that all of them overcome the baseline or are comparable with it, with the best performance obtained for 10~dB TNTSNR. These experiments could indicate that recorded soundscapes in \textbf{public} in general have a TNTSNR of about 10 -- 15~dB which should be confirmed by complementary experiments.

\section{EVALUATING ON NON-TARGET EVENTS ONLY}
\label{sec:notgeval}

Based on the previous experiments, TNTSNR could be one reason of mismatch between the synthetic soundscapes and the recorded soundscapes. But this could not explain all the performance differences observed here. In particular why in general having lower TNTSNR during training is decreasing the performance regardless of the validation. One possibility is that the system gets acoustically confused by a possible similarity in sound between events when soundscapes tend to be less dominated by target events. So we evaluated the system using the \textbf{synth\_ntg\_eval}, where only non-target events are considered, to see for which classes the system would output false positives. We evaluated the system on the \textbf{public} set; considering the systems trained for the first experiment (see Table~\ref{tab:results_public}). Results show that some sound events are detected more than others. For some classes as Speech, this could be explained by the original event distribution (indicated in the first column) but for some other classes as Dishes there is a discrepancy between the original distribution and the amount of false alarms. Interestingly the amount of false alarms is decreased sensibly for most of the classes when including non-target sound events during training. 

\begin{table}[!t]
\centering
 \begin{tabular}{l||c||c|c|c|c}
  \toprule
  &\multirow{2}{*}{Nref }&\multicolumn{4}{c}{Nsys}\\
  Classes& & A & B & C & Base \\
  \midrule
  Dog & 197 & 135 & 126 & 146 & 79 \\
  Vacuum\_cleaner & 127 & 31 & 42 & 44 & 47 \\
  Alarm\_bell & 191 & 47 & 50 & 52 & 59 \\
  Running\_water & 116 & 34 & 41 & 61 & 30 \\
  Dishes & 405 & 1478 & 395 & 1270 & 305 \\
  Blender & 100 & 63 & 32 & 55 & 19\\
  Frying & 156 & 70 & 41 & 60 & 33\\
  Speech & 1686 & 206 & 181 & 180 & 201 \\
  Cat & 141 & 99 & 103 & 98 & 73 \\
  Electric\_shaver & 103 & 21 & 18 & 18 & 7\\
 \bottomrule
 \end{tabular}
 \caption{Preliminary evaluation results by classes, evaluating the system with \textbf{synth\_ntg\_eval}. Nsys (A): training with \textbf{synth\_tg}, validating with \textbf{synth\_tg\_val}; Nsys (B): training with \textbf{synth\_tg\_ntg}, validating with \textbf{synth\_tg\_val}; Nsys (C): training with \textbf{synth\_tg}, validating with \textbf{synth\_tg\_ntg\_val}; Base: baseline using target and non-target events for training and validation.}
 \label{tab:results_last}
\end{table}  

\section{CONCLUSIONS AND FUTURE WORK}
\label{sec:conclusion}
This paper analyzes the impact of including non-target sound events in the synthetic soundscapes of the training dataset for SED systems trained on heterogeneous dataset. In particular, the experiments are divided into three stages: in the first part, we explore to what extend using non-target sound events at training has an impact on the system's performance, secondly we investigate the impact of varying TNTSNR and we conclude the study by analyzing a possible confusion of the SED model in case of false alarms triggered by non-target sound events. 

From the results reported on this paper, we can conclude that using non-target sound events can help the SED system to better detect the target sound events, but it is not clear to what extend and what would be the best way to generate the soundscapes. Results show that the final SED performance could depend on mismatches between synthetic and recorded soundscapes, part of which could be due to the  TNTSNR but not only. Results on the last experiment show that using non-target events at training decreases the amount of false alarms at test but from this experiment it is not possible to conclude on the impact of non-target sound events on the confusion between the target sound events. This is a first track for future investigation on the topic. Additionally, the impact of the non-target sound events at training on the ability of the system to better segment the target sound events in noisy soundscapes would have to be investigated. A final open question is the impact of the per class distribution of the sound events (both target and non-target) and their co-occurrence distribution on the SED performance.

\section{ACKNOWLEDGEMENTS}
\label{sec:acknowledgements}
We would like to thank all the other organizers of DCASE 2021 Challenge Task
4. In particular, we thank Eduardo Fonseca and Daniel P. W. Ellis for their help with the strong labels of the AudioSet dataset used to compute the events co-occurrences, and Justin Salamon and Prem Seetharaman for their help with Scaper.


\bibliographystyle{IEEEtran}
\bibliography{refs}

%
%
%
%
%
%
%
%
%

\end{sloppy}
\end{document}